\documentclass[doublecol]{epl2}
\usepackage{amssymb}

\newcommand{\la}{\langle}
\newcommand{\ra}{\rangle}
\newcommand{\lam}{\lambda}
\newcommand{\ti}{\tilde}
\newcommand{\ga}{\gamma}
\newcommand{\Ga}{\Gamma}
\newcommand{\ka}{\kappa}
\newcommand{\da}{\dagger}

\newcommand{\al}{\alpha}

\newcommand{\om}{\omega}

\newcommand{\pa}{\partial}
\newcommand{\non}{\nonumber}

\def\oc#1{{ Opt.\ Commun.} {\bf#1}}

\def\jpa#1{{ J.\ Phys.\ A} {\bf#1}}
\def\pra#1{{ Phys.\ Rev. A\/} {\bf#1}}
\def\prb#1{{ Phys.\ Rev. B\/} {\bf#1}}

\def\prl#1{{ Phys.\ Rev.\ Lett.} {\bf#1}}

\def\sci#1{{ Science} {\bf#1}}

\def\pla#1{{ Phys.\ Lett. A\/} {\bf#1}}
\def\rmp#1{{ Rev. \ Mod. \ Phys.} {\bf#1}}

\title{Stochastic Schr\"{o}dinger Equation for a Non-Markovian Dissipative
Qubit-Qutrit System}

\author{Jun Jing\inst{1,2}\footnote{Email address: Jun.Jing@stevens.edu} \and
Ting Yu\inst{1}\footnote{Email address: Ting.Yu@stevens.edu}}

\institute{
  \inst{1}Center for Controlled Quantum Systems, and the Department of
Physics and Engineering Physics, Stevens Institute of Technology, Hoboken, New
Jersey 07030, USA \\
  \inst{2}Department of Physics, Shanghai University, Shanghai 200444, China}


\abstract{We investigate the non-Markovian quantum dynamics of a hybrid open
system consisting of one qubit and one qutrit by employing a stochastic
Schr\"{o}dinger equation to generate diffusive quantum trajectories. We have
established an exact quantum state diffusion (QSD) equation for the dissipative
qubit-qutrit system coupled to a bosonic heat bath at zero temperature. As an
important application, the non-Markovian QSD equation is employed to simulate
the entanglement decay and generation measured by negativity. Finally, some
steady state properties of the hybrid system are also discussed.}

\begin{document}

\maketitle

\section{Introduction}\label{intro}

The recent development of quantum information and quantum optics
\cite{Nielsen,quantumoptics} has triggered much effort devoted to physics of
qubit systems in various physical settings as they are the most basic building
blocks of quantum computing and quantum information processing. The dynamical
behaviors of open quantum systems (qubit, qutrit, etc) are abundantly
documented for the cases where the environmental noise is weak and the
bandwidth of the noise is broad such that the so-called Born-Markov
approximation gives rise to a reliable description of the dynamics of the
quantum open system \cite{Gardiner}. Under this approximation, one may employ
either the standard Lindblad master equations or their alternatives including
quantum trajectories, Fokker-Planck equations and Langevin equations, to name
a few \cite{Dalibard,Gardiner2,Carmicheal,Gisin-Percival}. When the external
environment is a structured medium or the system-environment coupling is
strong, then the non-Markovian equations of the motion such as non-Markovian
master equations have to be used to describe the time evolution of the reduced
density operator of the system of interest
\cite{Breuer,NonMarkov1,NonMarkov2,NonMarkov3,NonMarkov4}.

A stochastic Schr\"{o}dinger equation named non-Markovian quantum state
diffusion equation by Strunz and his coworkers has been applied to many
interesting physical systems including a dissipative two-level system, quantum
Brownian motion model and a multiple qubit system
\cite{Diosi97,Diosi,Strunz99,Yupra99,Strunz01CP,Wiseman,Strunz04PRA,Yu04PRA,
Vega,Bassi,Jingetal10}.
A recent application of the non-Markovian QSD to molecular aggregates can be
found in \cite{StrunzPRL}. It has been known that the higher dimensional
quantum systems (qutrit, qudit, etc.) have demonstrated potential applications
in quantum information processing \cite{Khan,Checinska}. Their dynamical aspect
in a non-Markovian regime is still not well understood due to the lack of exact
non-Markovian QSD equation or exact master equation. In the framework of
quantum open system, a qubit-qutrit system is clearly the first non-trivial
extension beyond the qubit-qubit system, yet it is sophisticated enough to
provide useful insight into non-Markovian dynamics of high dimensional open
systems \cite{Strunz99,Jing_Yu10}. In the case of entanglement dynamics, the
qubit-qutrit system allows a rigorous characterization of entanglement when the
negativity is employed. It should be noted that the disentanglement pathways
for a qubit-qutrit system have been studied when the qubit and qutrit are
coupled to individual Markov environments, respectively \cite{Ann}, but a
systematic investigation into the qubit-qubtrit model in a non-Markovian regime
is still missing.

Purpose of this Letter is to generalize the non-Markovian QSD approach to the
case of the qubit-qutrit system. In particular, we have derived the exact
time-local quantum state diffusion equation for the qubit-qutrit system, in
which the explicit formation of O-operator is very different from that in the
two-qubit case \cite{XJBT}. In what follows, we will first introduce the
general formalism of non-Markovian quantum trajectory. We then derive the exact
time-local QSD equation for the qubit-qutrit dissipative model at zero
temperature. Finally, by employing the non-Markovian QSD equation, we
investigate the negativity dynamics of the hybrid system for different
anisotropic coupling parameters, environmental memories and initial states. The
details for the exact QSD equation are left to the two appendices.

\section{General formalism}\label{NMQT}

The model considered in this paper is described by the qubit-qutrit system
coupled linearly to a general bosonic environment consisting of a set of
harmonic oscillators:
\begin{equation}\label{Htot}
H_{\rm tot}=H_{\rm sys}+\sum_{\lam}(g_{\lam}La_{\lam}^\da+g_{\lam}^*L^\da
a_{\lam})+\sum_{\lam}\omega_{\lam}a_{\lam}^\da a_{\lam},
\end{equation}
where $H_{\rm sys}=\omega_AS_z^A+\omega_BS_z^B$ and $L=S_-^A+\ka S_-^B$. $\ka$
depicts the asymmetry between the coupling coefficients of the two subsystems
with the environment, and explicitly the notations are given by
\begin{eqnarray}\non
S_z^A&=&\left(\begin{array}{ccc}
      1 & 0 & 0  \\
      0 & 0 & 0 \\
      0 & 0 & -1
    \end{array}\right), \quad
S_z^B=\frac{1}{2}\left(\begin{array}{cc}
      1 & 0  \\
      0 & -1
    \end{array}\right), \\
S_-^A&=&\left(\begin{array}{ccc}
      0 & 0 & 0  \\
      1 & 0 & 0 \\
      0 & 1 & 0
    \end{array}\right), \quad
S_-^B=\left(\begin{array}{cc}
      0 & 0  \\
      1 & 0
    \end{array}\right).
\end{eqnarray}

In what follows we will show that a time-local stochastic Schr\"{o}dinger
equation at zero temperature called QSD equation can be derived from the above
microscopic model. Using the Bargmann state bases for the environment degrees
of freedom, the total quantum state including the system and the environment
could be written as
\begin{equation}
|\Psi_{\rm tot}(t)\ra=\int\left[\frac{d^2z}{\pi}\right]e^{-|z|^2}|\psi_t(z)\ra
||z\ra,
\end{equation}
where $||z\ra=||z_1\ra||z_2\ra\cdots||z_\lam\ra$ with $||z_\lam\ra=e^{z_\lam
a^\da_\lam}|0\ra$, $|z|^2=\sum_\lam|z_\lam|^2$, and
$[\frac{d^2z}{\pi}]=\frac{d^2z_1}{\pi}\frac{d^2z_2}{\pi}\cdots
\frac{d^2z_\lam}{\pi}\cdots$ . It was shown that the wave function of the
central system $\psi_t(z)$ is governed by a linear stochastic differential
equation with a functional derivative of the pure state of system over the
stochastic noise $z$ \cite{Diosi,Strunz99}:
\begin{equation}\label{LQSD}
\frac{\partial}{\partial t}\psi_t(z)
=\left[-iH_{\rm sys}+Lz_t-L^\da\int_0^tds\alpha(t,s)\frac
{\delta}{\delta z_s}\right]\psi_t(z),
\end{equation}
where $\alpha(t,s)$ is the environmental correlation function and
$z_t=-i\sum_\lam g_\lam z^*_\lam e^{i\omega_\lam t}$ is a complex Gaussian
process satisfying $M[z_t]=M[z_tz_s]=0$ and $M[z^*_tz_s]=\alpha(t,s)$. The
density matrix of the system can be recovered by the ensemble average of
quantum trajectories:
\begin{equation}\label{rhot}
\rho_t=M[|\psi_t(z)\ra\la\psi_t(z)|]=\int\frac{d^2z}{\pi}e^{-|z|^2}
|\psi_t(z)\ra\la\psi_t(z)|.
\end{equation}
One of the advantages of the non-Markovian quantum trajectory method is that a
small number of trajectories could also be used to estimate qualitatively the
dynamical behaviors of the open system \cite{Jing_Yu10}. Crucial to application
of the general QSD equation (\ref{LQSD}) to a practical problem is to transform
eq.~(\ref{LQSD}) into a time-local stochastic differential equation:
$\frac{\delta\psi_t(z)}{\delta z_s}=O(t,s,z)\psi_t(z)$. Then we have
\begin{equation}\label{LQSDe}
\frac{\partial}{\partial
t}\psi_t(z)=[-iH_{\rm sys}+Lz_t-L^\da\bar{O}(t,z)]\psi_t(z),
\end{equation}
where $\bar{O}(t,z)$ is defined as $\int_0^t\alpha(t,s)O(t,s,z)ds$. Although
there is no general recipe to construct the O-operator explicitly, the exact
O-operator for this qubit-qutrit dissipative model can be constructed by the
following expression:
\begin{eqnarray}\non
O(t,s,z)&=&\sum_{j=1}^8f_j(t,s)D_j+\sum_{j=9}^{12}\int_0^tp_j(t,s,s_1)z_{s_1}
ds_1D_j \\  \label{Oop} &+&
\int_0^t\int_0^tp_{13}(t,s,s_1,s_2)z_{s_1}z_{s_2}ds_1ds_2D_{13},
\end{eqnarray}
where the operators $D_j$, $j=1,2,\cdots,13$, are given by
\begin{eqnarray*}
D_1&=&|0\ra_A\la2|\otimes\sigma^B_+, \quad
D_2=|1\ra_A\la2|\otimes|1\ra_B\la1|, \\
D_3&=&|0\ra_A\la1|\otimes|1\ra_B\la1|, \quad
D_4=|1\ra_A\la2|\otimes|0\ra_B\la0|, \\
D_5&=&|0\ra_A\la1|\otimes|0\ra_B\la0|, \quad
D_6=|2\ra_A\la2|\otimes\sigma^B_-, \\
D_7&=&|1\ra_A\la1|\otimes\sigma^B_-, \quad D_8=|0\ra_A\la0|\otimes\sigma^B_-,\\
D_9&=&|0\ra_A\la2|\otimes|1\ra_B\la1|, \quad
D_{10}=|0\ra_A\la2|\otimes|0\ra_B\la0|, \\
D_{11}&=&|1\ra_A\la2|\otimes\sigma^B_-, \quad
D_{12}=|0\ra_A\la1|\otimes\sigma^B_-, \\
D_{13}&=&|0\ra_A\la2|\otimes\sigma^B_-,
\end{eqnarray*}
where $\sigma_{\pm}^B$ are the Pauli matrices. And the coefficient functions
$f_j$'s, $j=1,2,\cdots, 8$ and $p_j$'s, $j=9,\cdots, 13$ could be determined by
the consistency condition \cite{Diosi97}: $\frac{\delta}{\delta
z_s}\frac{\partial\psi_t}{\partial t}=\frac{\partial}{\partial
t}\frac{\delta\psi_t}{\delta z_s}$. It turns out to be
\begin{equation}\label{LQSD2}
\frac{\partial O(t,s,z)}{\partial t}=\left[-iH_{\rm sys}+Lz_t
-L^\da\bar{O},O\right]-L^\da\frac{\delta\bar{O}}{\delta z_s}.
\end{equation}
As a primary result in this paper, we have explicitly derived the O-operator
for the qubit-qutrit model, therefore we have established for the first time
the exact dynamical equation for this hybrid system. From the initial condition
for the O-operator $O(s,s,z)=L$, we have $f_1(s,s)=0$; $f_j(s,s)=1$,
$j=2,3,4,5$; $f_j(s,s)=\ka$, $j=6,7,8$; $p_j(s,s,s_1)=0$, $j=9,10,11,12$; and
$p_{13}(s,s,s_1,s_2)=0$. The partial differential equations and the boundary
conditions for $f_j$'s and $p_j$'s are given in Appendix A.

The linear QSD equation in eq.~(\ref{LQSDe}) does not conserve the norm of the
state vector $\psi_t(z)$, so for numerical simulations, one typically employs
the nonlinear version of the QSD equation (\ref{LQSDe}) \cite{Strunz99}:
\begin{eqnarray}\non
\frac{d}{dt}\ti{\psi}_t(z)&=&[-iH_{\rm sys}
+\Delta_t(L)\ti{z}+\Delta_t(L^\da\bar{O}(t,\ti{z})) \\ \label{nLQSD}
&-&\la L^\da\ra_t\Delta_t(\bar{O}(t,\ti{z}))]\ti{\psi}_t(z),
\end{eqnarray}
where the norm-preserved wave function is defined as
$\ti{\psi}_t(z)=\frac{\psi_t(z)}{||\psi_t(z)||}$. Note that
$\ti{z}_t=z_t+\int_0^t\alpha^*(t,s)\la L^\da\ra_sds$ is the shift noise,
$\Delta_t(Q)\equiv Q-\la Q\ra_t$ for any operator $Q$, and $\la
Q\ra_t\equiv\la\psi_t|Q|\psi_t\ra$ denotes the quantum average. Explicitly, the
operator $\bar{O}(t,\ti{z})$ may be written as
\begin{eqnarray}\non
\bar{O}(t,\ti{z})&=&\sum_{j=1}^8F_j(t)D_j+\sum_{j=9}^{12}\int_0^tP_j(t,s_1)
\ti{z}_{s_1}ds_1D_j \\  \label{Oop2} &+&
\int_0^t\int_0^tP_{13}(t,s_1,s_2)\ti{z}_{s_1}\ti{z}_{s_2}ds_1ds_2D_{13},
\end{eqnarray}
where the definition of $F_j(t)$'s, $j=1,\cdots, 8$, $P_j(t,s_1)$'s,
$j=9,\cdots, 12$, and $P_{13}(t,s_1,s_2)$ can be found in Appendix A.

\section{Simulations and Discussions}\label{Simulation}

\begin{figure}[htbp]
\centering
\includegraphics[width=2.6in]{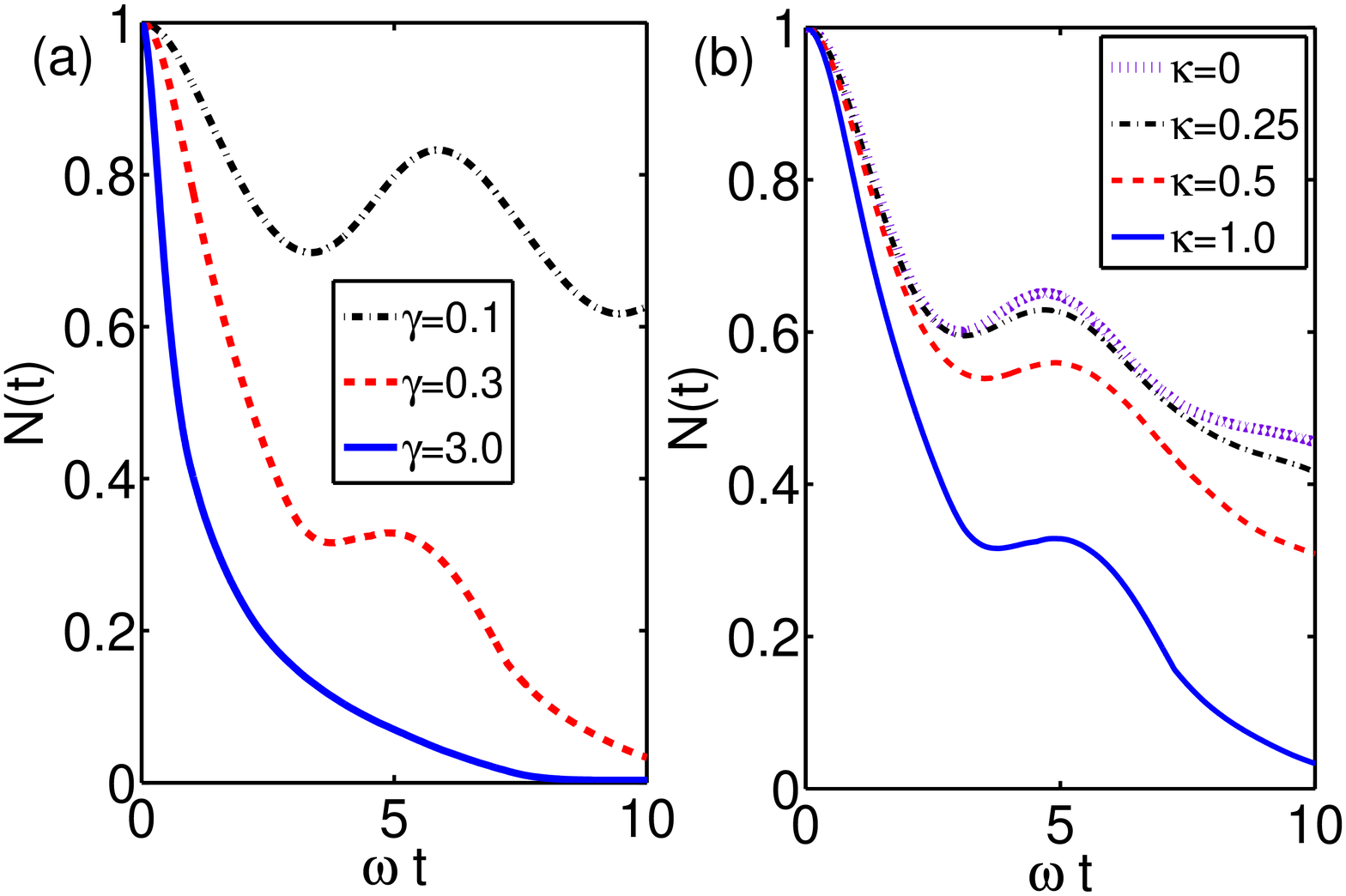}
\caption{The negativity as function of $\om t$ with (a) different $\ga$'s
($\ka=1$), and (b) different $\kappa$'s ($\ga=0.3$) obtained by an average over
$1000$ trajectories. The initial state is chosen as
$(1/\sqrt{2})(|1\ra_A|1\ra_B+|0\ra_A|0\ra_B)$.}\label{Nga0011}
\end{figure}

\begin{figure}[htbp]
\centering
\includegraphics[width=2.6in]{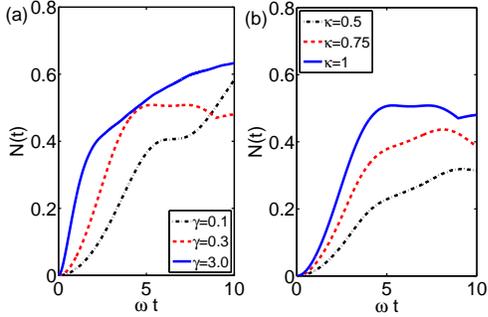}
\caption{The negativity as function of $\om t$ with (a) different $\ga$'s
($\ka=1$), and (b) different $\kappa$'s ($\ga=0.3$) obtained by an average over
$1000$ trajectories. The initial state is chosen as $|2\ra_A|0\ra_B$.}
\label{Nga21}
\end{figure}

Entanglement dynamics of the qubit-qubit system in Markov and non-Markovian
baths has been extensively studied in the last decade
\cite{esdprl1,Yu-Eberly2002,esdprl2,Buchleitner04,Bellomo,Sabrina,Ficek,Zhupra,
SunCP,Strunz09,Yu-Eberly10}. Equipped with the exact QSD equation for the
qubit-qutrit system, we are in the position to study rigorously the
non-Markovian entanglement and decoherence dynamics for this hybrid system. For
the purpose of entanglement dynamics, qubit-qubit ($2\otimes2$) and
qubit-qutrit ($2\otimes3$) systems are particularly interesting due to the fact
that entanglement measured negativity \cite{Peres,Horodeckipla,Vidal,Horodecki}
serves both sufficient and necessary conditions for an arbitrary state of these
two bipartite systems. Precisely, the negativity is defined as
$N(\rho)\equiv||\rho^{T_B}||-1=2\sum_j|\mu_j|$, where $T_B$ stands for the
partial transpose with respect to one of subsystems $B$, and $\mu_j$'s are the
negative eigenvalues of the matrix $\rho^{T_B}$. A qubit-qutrit state
represented by the density matrix $\rho$ is entangled if and only if $N(\rho)$
is positive. The negativity $N(\rho)$ varies from zero for all separable states
to unity for the maximally entangled states. For a general pure state of
qubit-qubit and qubit-qutrit system, it reduces to the definition of Wootters'
concurrence and generalized concurrence for pure states \cite{Wootters,Fei},
respectively.

Once the correlation function $\alpha(t,s)$ is chosen, it is straightforward to
calculate these coefficient functions in eq.~(\ref{Oop2}). In the following,
for the sake of simplicity, we consider a Gaussian complex noise $z_t$
satisfying the Ornstein-Uhlenbeck process:
$\alpha(t,s)=\frac{\Gamma\ga}{2}e^{-\ga|t-s|}$, where $\Gamma$ is the
dissipation rate and $\gamma$ describes the memory time of the environment.
More precisely, $\gamma^{-1}$ defines the finite correction time of the
environment. When $\gamma\rightarrow\infty$, the correlation function
approaches the Markov limit: $\al(t,s)=\Gamma\delta(t-s)$. In Appendix B, the
details of the differential equations for all the functions $F_j$'s
($j=1,\cdots,8$) and $P_j$'s ($j=9,\cdots,13$) are provided. It should be noted
that the $P_j$ terms contain explicitly the history integral over complex
Gaussian noise $z_t$. It is expected that they will be important in
non-Markovian dynamics when $\ga$ is not too large (far from Markov regimes).

In the following numerical simulations, we choose $\om_A=\om_B=\om$, $\Gamma=1$
and the double integral term about $P_{13}$ containing noises is neglected to
save the computer memory. It is easy to see that the higher orders are clearly
less important in a weakly non-Markovian regime.

In figs.~\ref{Nga0011}(a) and \ref{Nga0011}(b), the initial state is a maximal
entangled state $|\psi(0)\rangle=(1/\sqrt{2})(|1\ra_A|1\ra_B+|0\ra_A|0\ra_B)$.
Fig.~\ref{Nga0011}(a) shows the effect of $\ga$ on the decay rates of the
entanglement of the system. When $\ga$ is as large as $3$, the dynamics is seen
to be very close to the Markov limit. It shows the reservoir can quickly ruin
the coherence of the initial state, which evolves into a separable state
monotonously. When $\ga=0.1$, the environment has a long memory time, so that
the non-Markovian features become dominant, hence the negativity exhibits a
strong fluctuation pattern and maintains a highly entangled state for a long
time. When $\ga=0.3$, it shows a moderate non-Markovian behavior. 
Fig.~\ref{Nga0011}(b) illustrates how the anisotropic coupling parameter $\ka$
affects the entanglement evolution. The asymmetrical coupling slows down the
entanglement decay gradually as the parameter $\kappa$ becomes smaller and
smaller. The extreme case is that $\ka=0$, where one of the subsystems is
completely decoupled from the influence of the environment.

Figure.~\ref{Nga21}(a) plots the entanglement generation of the separable state
$|\psi(0)\rangle=|2\ra_A|0\ra_B$. If the qubit is not interacting with the
qutrit, then the entanglement generation is purely due to the indirect
interaction between the subsystems induced by the common bath and memory times.
For this initially separable state, the larger $\ga$ (near-Markov reservoir)
leads to a faster transition from $|2\ra$ to $|1\ra$ compared to the smaller
$\gamma$ (a stronger non-Markovian regime). Interestingly, the near-Markov
reservoir in turn causes a quicker generation of entanglement with one
excitation transition in a short time interval. For some initially separable
states, the non-Markovian property may be essential in entanglement generation.
In the case of $\ga=0.1$, representing a highly non-Markovian regime, it
clearly shows that the dynamics generates a higher degree of entanglement than
that in the case of $\ga=0.3$ after $\om t\approx9$. While fig.~\ref{Nga21}(b)
describes the effect of different $\ka$'s on the entanglement generation with
the same $\ga$. It is very interesting to see that the entangling power is
highly sensitive to the coupling balance for the qubit and qutrit interacting
with the common bath.

\begin{figure}[htbp]
\centering
\includegraphics[width=2.6in]{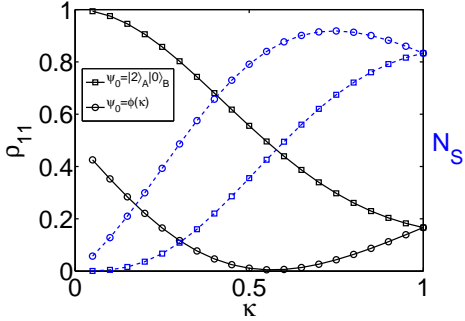}
\caption{The distribution of the ground state $\rho_{11}$ (black solid lines)
and the negativity of the steady state $N_s$ (blue dashed lines) as functions
of $\ka$ with different initial states in the long-time limit.} \label{DNS}
\end{figure}

It is interesting to examine the long-time behavior of the combined
qubit-qutrit system under the influence of the non-Markovian environment. For
the Ornstein-Uhlenbeck noise, the information for the long-time limit may be
obtained by solving the Markov master equation for the qubit-qutrit system
\begin{equation}\label{MME}
\partial_t\rho_t=-i[H_{\rm sys},\rho_t]+\frac{\Ga}{2}[L,\rho_tL^\da]
+\frac{\Ga}{2}[L\rho_t,L^\da].
\end{equation}
Now we consider several interesting initial states in the four subspaces
spanned by (i) $|2\ra_A|1\ra_B$, (ii) $|2\ra_A|0\ra_B$ and $|1\ra_A|1\ra_B$,
(iii) $|1\ra_A|0\ra_B$ and $|0\ra_A|1\ra_B$, (iv) $|0\ra_A|0\ra_B$,
respectively.  When $\Ga=1$, $\om_A=\om_B$,  we have
\begin{eqnarray*}
\pa_t\rho_{66}&=&-(1+\ka^2)\rho_{66}, \\
\pa_t\rho_{55}&=&\ka^2\rho_{66}-\frac{\ka}{2}(\rho_{45}+\rho_{54})-\rho_{55},\\
\pa_t\rho_{54}&=&\ka\rho_{66}-\frac{\ka}{2}(\rho_{44}+\rho_{55})
-(1+\frac{\ka^2}{2})\rho_{54}, \\
\pa_t\rho_{44}&=&\rho_{66}-(1+\ka^2)\rho_{44}
-\frac{\ka}{2}(\rho_{54}+\rho_{45}), \\
\pa_t\rho_{33}&=&\rho_{55}-\rho_{33}-\frac{\ka}{2}(\rho_{23}+\rho_{32})
+\ka(\rho_{45}+\rho_{54})+\ka^2\rho_{44}, \\
\pa_t\rho_{32}&=&\rho_{54}-(\frac{1}{2}+\frac{\ka^2}{2})\rho_{32}
-\frac{\ka}{2}(\rho_{22}+\rho_{33})+\ka\rho_{44}, \\
\pa_t\rho_{22}&=&\rho_{44}-\frac{\ka}{2}(\rho_{32}+\rho_{23})-\ka^2\rho_{22},\\
\pa_t\rho_{11}&=&\rho_{33}+\ka(\rho_{23}+\rho_{32})+\ka^2\rho_{22},
\end{eqnarray*}
where the notations for the six basis vectors are $|6\ra=|2\ra_A|1\ra_B$,
$|5\ra=|2\ra_A|0\ra_B$, $|4\ra=|1\ra_A|1\ra_B$, $||3\ra=|1\ra_A|0\ra_B$,
$|2\ra=|0\ra_A|1\ra_B$, and $|1\ra=|0\ra_A|0\ra_B$. From the above equations,
one can easily conclude that for the stationary solutions we always have
$\rho_{66}=\rho_{55}=\rho_{54}=\rho_{44}=0$ (they are independent on the
initial states). But $\rho_{33}=-\ka\rho_{32}$, $\rho_{22}=-\rho_{32}/\ka$,
$\rho_{11}=1-\rho_{33}-\rho_{22}$, which are dependent on the initial states.
Therefore, we have
\begin{eqnarray}\label{steady}
\rho_{t=\infty}&=&\rho_{11}(\infty)|00\ra_{AB}\la 00| \non \\
&+&[1-\rho_{11}(\infty)]|\psi(\ka)\ra\la\psi(\ka)|,
\end{eqnarray}
where $|\psi(\ka)\rangle=(1/\sqrt{1+\ka^2})(\ka|1\ra_A|0\ra_B-|0\ra_A|1\ra_B)$,
which is an eigenstate of the Lindblad operator $L=S_-^A+\ka S_-^B$.

The final value of $\rho_{11}$ is a function of the $\ka$ and the initial
states as well as the negativity of the final state. The black-solid line with
the square marker in Fig.~(\ref{DNS}) describes the probability distribution
$\rho_{11}$ of the ground state $|0\ra_A|0\ra_B$ for the initial state
$|\psi(0)\ra=|2\ra_A|0\ra_B$. In the range $0<\ka\leqslant1$, it decreases with
$\ka$ monotonously. And the negativity [equals to
$2(1-\rho_{11})\ka/(1+\ka^2)$] is shown by the blue-dashed line with the same
marker. It coincides with the curves in Fig.~\ref{Nga21}(b). We also plot the
numerical results (the two lines with the circle marker) for the initial
state $|\phi(\ka)\rangle=(1/\sqrt{1+\ka^2})(\ka|2\ra_A|0\ra_B-|1\ra_A|1\ra_B)$,
which is also an eigenstate of $L$, but this one is not protected by the common
dissipative environment. It shows clearly that the state evolves into a mixture
of the ground state $|0\ra_A|0\ra_B$ and $|\psi(\ka)\rangle$ given by 
(\ref{steady}). An interesting result arising from the above discussions is
that the residue entanglement degree reaches its maximal value at $\ka \approx
0.75$ instead of $1$ which corresponds to the balanced coupling. It is also
interesting to see that the final state could become an approximate pure state
when $\ka \approx 0.55$, which demonstrates a rather different patterns from
those initially separable states considered in this section.

\section{Conclusion}\label{Conclusion}

In summery, we have investigated non-Markovian QSD approach to the qubit-qutrit
system coupled to a zero-temperature bath. With the non-Markovian QSD equation,
the entanglement dynamics of the qubit-qutrit system is studied carefully. The
numerical simulations have been used to show the negativity dynamics for
different initial states. We also discuss the long-time behaviors of the
initially entangled or separable states of the hybrid system. Entanglement
generation and decoherence have been shown to be sensitively dependent on
coupling strength, memory time and initial states. The approach described here
provides a fundamental tool for qubit-qutrit systems when the non-Markovian
properties become important.

\acknowledgments
This work has benefited from the interesting discussions with Profs. J. H.
Eberly and B. L. Hu. We acknowledge the support by grants from DARPA QuEST
No. HR0011-09-1-0008, the NSF PHY-0925174, and the NSFC 10804069.

\section{\textbf{Appendix A: Equations of the Motions for the Coefficients and
Boundary Conditions - General Case}}\label{appa}

\setcounter{equation}{0}
\renewcommand{\theequation}{A\arabic{equation}}

By the time evolution equation for O-operator in eq.~(\ref{LQSD2}), we get a
set of partial differential equations about the coefficient functions in
eq.~(\ref{Oop}):
\begin{eqnarray}\non
\partial_tf_1(t,s)&=&i(2\omega_A-\omega_B)f_1+F_1f_3+F_4f_1+\ka F_4f_3
\\ &-& \ka F_5f_4-\ka F_8f_1-\ka P_{10}(t,s),
\end{eqnarray}
\begin{eqnarray}\non
\partial_tf_2(t,s)=i\omega_Af_2+F_2f_2-F_1f_6-F_3f_2-\ka F_4f_6\\
+\ka F_6f_2-\ka F_7f_2-P_9(t,s)-\ka P_{11}(t,s),
\end{eqnarray}
\begin{eqnarray}\non
\partial_tf_3(t,s)&=&i\omega_Af_3+F_3f_3-\ka F_5f_7+F_7f_1+\ka
F_7f_3\\ &-&\ka F_8f_3-\ka P_{12}(t,s),
\end{eqnarray}
\begin{eqnarray}\non
\partial_tf_4(t,s)&=&i\omega_Af_4+F_1f_7+F_4f_4+F_4f_7-F_5f_4
\\ &-&F_8f_1-P_{10}(t,s),
\end{eqnarray}
\begin{equation}
\partial_tf_5(t,s)=i\omega_Af_5+F_5f_5+\ka F_5f_8,
\end{equation}
\begin{eqnarray}\non
\partial_tf_6(t,s)&=& i\omega_Bf_6+F_2f_6-F_4f_6+\ka F_6f_6\\
&-&F_7f_2-P_{11}(t,s),
\end{eqnarray}
\begin{eqnarray}\non
\partial_tf_7(t,s)&=&i\omega_Bf_7+F_3f_7-F_5f_7+F_7f_4+\ka F_7f_7\\ &-&F_8f_3
-P_{12}(t,s),
\end{eqnarray}
\begin{equation}
\partial_tf_8(t,s)=i\omega_Bf_8+F_8f_5+\ka F_8f_8,
\end{equation}
\begin{eqnarray}\non
\partial_tp_9(t,s,s_1)=2i\omega_Ap_9+F_2p_9-\ka F_5p_{11}+\ka F_6p_9 \\ \non
-\ka F_8p_9+P_9f_3-\ka P_{10}f_6+P_{11}f_1 \\
+\ka P_{11}f_3-\ka P_{12}f_2-2\ka P_{13}(t,s,s_1),
\end{eqnarray}
\begin{eqnarray}\non
\partial_tp_{10}(t,s,s_1)&=&2i\omega_Ap_{10}+F_1p_{12}+F_4p_{10}+\ka F_4p_{12}
\\ &+&P_{10}f_5+\ka P_{10}f_8,
\end{eqnarray}
\begin{eqnarray}\non
\partial_tp_{11}(t,s,s_1)&=&i(\omega_A+\omega_B)p_{11}+F_2p_{11}-F_5p_{11}
\\ \non &+& \ka F_6p_{11}-F_8p_9+P_9f_7-P_{10}f_6+P_{11}f_4 \\
&+&\ka P_{11}f_7-P_{12}f_2-2P_{13}(t,s,s_1),
\end{eqnarray}
\begin{eqnarray}\non
\partial_tp_{12}(t,s,s_1)&=&i(\omega_A+\omega_B)p_{12}+F_3p_{12}+F_7p_{10}
\\ &+& \ka F_7p_{12}+P_{12}f_5+\ka P_{12}f_8,
\end{eqnarray}
\begin{eqnarray}\non
\partial_tp_{13}(t,s,s_1,s_2)&=&i(2\omega_A+\omega_B)p_{13}+F_2p_{13}
+\ka F_6p_{13}\\ \non
&+&P_9p_{12}+P_{11}p_{10}+\ka P_{11}p_{12}+P_{13}f_5\\ &+&\ka P_{13}f_8,
\end{eqnarray}
where $F_j=F_j(t)\equiv\int_0^t\alpha(t,s)f_j(t,s)ds$, $j=1,\cdots,8$;
$P_j=P_j(t,s_1)\equiv\int_0^t\alpha(t,s)p(t,s,s_1)ds$, $j=9,\cdots,12$;
$P_{13}=P_{13}(t,s_1,s_2)\equiv\int_0^t\alpha(t,s)p_{13}(t,s,s_1,s_2)ds$. And
the boundary conditions are:
\begin{eqnarray}\non
p_9(t,s,t)&=&-\ka f_1(t,s)+f_2(t,s)-f_3(t,s),  \\ \non
p_{10}(t,s,t)&=&\ka f_1(t,s)+f_4(t,s)-f_5(t,s), \\ \non
p_{11}(t,s,t)&=&\ka f_2(t,s)-\ka f_4(t,s)+f_6(t,s)-f_7(t,s), \\ \non
p_{12}(t,s,t)&=&\ka f_3(t,s)-\ka f_5(t,s)+f_7(t,s)-f_8(t,s), \\ \non
p_{13}(t,s,t,s_1)&=&0.5[\ka p_9(t,s,s_1)-\ka p_{10}(t,s,s_1) \\
\label{boundary} &+&p_{11}(t,s,s_1)-p_{12}(t,s,s_1)].
\end{eqnarray}

\section{\textbf{Appendix B: Equations of the Motion for the Coefficients:
Ornstein-Uhlenbeck Noise}}\label{appb}

\setcounter{equation}{0}
\renewcommand{\theequation}{B\arabic{equation}}

In the case of the Ornstein-Uhlenbeck process, by the definitions and the
boundary conditions in eq.~(\ref{boundary}), we get a set of closed
differential equations for $F_j$'s ($j=1,\cdots,8$), $\bar{P}_j$'s
($j=9,\cdots,12$) and $\ti{P}_{13}$:
\begin{eqnarray}\non
\partial_tF_1(t)&=&(-\gamma+2i\omega_A-i\omega_B)F_1+
F_1F_3+F_1F_4 \\
&-&\ka F_1F_8+\ka F_3F_4-\ka F_4F_5-\ka\bar{P}_{10},
\end{eqnarray}
\begin{eqnarray}\non
\partial_tF_2(t)&=&\frac{\Ga\gamma}{2}+(-\gamma+i\omega_A)F_2+
F_2^2-F_1F_6-F_2F_3\\
&+&\ka F_2F_6 -\ka F_2F_7-\ka F_4F_6-\bar{P}_9-\ka\bar{P}_{11},
\end{eqnarray}
\begin{eqnarray}\non
\partial_tF_3(t)&=&\frac{\Ga\gamma}{2}+(-\gamma+i\omega_A)F_3+
F_1F_7+F_3^2+\ka F_3F_7 \\
&-&\ka F_3F_8-\ka F_5F_7-\ka\bar{P}_{12},
\end{eqnarray}
\begin{eqnarray}\non
\partial_tF_4(t)&=&\frac{\Ga\gamma}{2}
+(-\gamma+i\omega_A)F_4+F_1F_7-F_1F_8+F_4^2\\ &-&F_4F_5+F_4F_7-\bar{P}_{10},
\end{eqnarray}
\begin{equation}
\partial_tF_5(t)=\frac{\Ga\gamma}{2}+(-\gamma+i\omega_A)F_5+F_5^2+\ka F_5F_8,
\end{equation}
\begin{eqnarray}\non
\partial_tF_6(t)&=&\frac{\ka\Ga\gamma}{2}+(-\gamma+i\omega_B)F_6
+F_2F_6-F_2F_7 \\ &-& F_4F_6+\ka F_6^2-\bar{P}_{11},
\end{eqnarray}
\begin{eqnarray}\non
\partial_tF_7(t)&=&\frac{\ka\Ga\gamma}{2}+(-\gamma+i\omega_B)F_7
+F_3F_7-F_3F_8\\ &+& F_4F_7-F_5F_7+\ka F_7^2-\bar{P}_{12},
\end{eqnarray}
\begin{equation}
\partial_tF_8(t)=\frac{\ka\Ga\gamma}{2}+(-\gamma+i\omega_B)F_8+F_5F_8+\ka F_8^2,
\end{equation}
\begin{eqnarray}\non
\partial_t\bar{P}_9&=&(-2\gamma+2i\omega_A)\bar{P}_9+
\frac{\Ga\gamma}{2}(-\ka F_1+F_2-F_3)+F_1\bar{P}_{11} \\ \non
&+&F_2\bar{P}_9-\ka F_2\bar{P}_{12}+F_3\bar{P}_9+\ka F_3\bar{P}_{11}-\ka
F_5\bar{P}_{11} \\ &+&\ka F_6\bar{P}_9-\ka F_6\bar{P}_{10}-\ka
F_8\bar{P}_9-2\ka\ti{P}_{13},
\end{eqnarray}
\begin{eqnarray}\non
\partial_t\bar{P}_{10}&=&(-2\gamma+2i\omega_A)\bar{P}_{10}+
\frac{\Ga\gamma}{2}(\ka F_1+F_4-F_5)+F_1\bar{P}_{12} \\
&+&F_4\bar{P}_{10}+\ka F_4\bar{P}_{12}+F_5\bar{P}_{10}+\ka F_8\bar{P}_{10},
\end{eqnarray}
\begin{eqnarray}\non
\partial_t\bar{P}_{11}&=&(-2\gamma+i\omega_A+i\omega_B)\bar{P}_{11}+
\frac{\Ga\gamma}{2}(\ka F_2-\ka F_4+F_6\\ \non &-&F_7)
+F_2\bar{P}_{11}-F_2\bar{P}_{12}+F_4\bar{P}_{11}-F_5\bar{P}_{11}
-F_6\bar{P}_{10} \\ &+&\ka F_6\bar{P}_{11}
+F_7\bar{P}_9+\ka F_7\bar{P}_{11}-F_8\bar{P}_9-2\ti{P}_{13},
\end{eqnarray}
\begin{eqnarray}\non
\partial_t\bar{P}_{12}&=&(-2\gamma+i\omega_A+i\omega_B)\bar{P}_{12}+
\frac{\Ga\gamma}{2}(\ka F_3-\ka F_5+F_7 \\ \non &-&F_8)
+F_3\bar{P}_{12}+F_5\bar{P}_{12}+F_7\bar{P}_{10} \\
&+&\ka F_7\bar{P}_{12}+\ka F_8\bar{P}_{12},
\end{eqnarray}
\begin{eqnarray}\non
\partial_t\ti{P}_{13}&=&(-3\gamma+2i\omega_A+i\omega_B)\ti{P}_{13}
+\frac{\Ga\gamma}{4}(\ka\bar{P}_9-\ka \bar{P}_{10} \\ \non
&+& \bar{P}_{11}-\bar{P}_{12})+F_2\ti{P}_{13}+F_5\ti{P}_{13}+\ka
F_6\ti{P}_{13} \\ &+&\bar{P_9}\bar{P}_{12}+
\ka F_8\ti{P}_{13}+\bar{P}_{10}\bar{P}_{11}+\ka\bar{P}_{11}\bar{P}_{12},
\end{eqnarray}
where $\bar{P}_j=\bar{P}_j(t)\equiv\int_0^t\alpha(t,s)P_i(t,s)ds$,
$j=9,\cdots,12$;
$\tilde{P}_{13}=\int_0^t\int_0^t\alpha(t,s_1)\alpha(t,s)P_{13}(t,s,s_1)dsds_1$;
And the initial conditions for $F_j(0), j=1,\cdots,8$, $\bar{P}_j(0),
j=9,\cdots,12$ and $\ti{P}_{13}(0)$ are all zero.

\end{document}